\documentclass[conference]{IEEEtran}
\IEEEoverridecommandlockouts
\usepackage{cite}
\usepackage[numbers,sort&compress]{natbib}

\usepackage[utf8]{inputenc}
\usepackage{amsmath,amssymb}
\usepackage{multirow}
\usepackage{graphicx}
\usepackage{bm}
\usepackage{enumerate}
\usepackage{comment}
\usepackage{xcolor}
\usepackage{url}
\usepackage{color,soul}
\usepackage{subcaption}
\usepackage{float}
\usepackage{booktabs} % For prettier tables
\usepackage{csquotes}
\definecolor{light-gray}{gray}{0.8}

\usepackage{gensymb}
\let\labelindent\relax
\usepackage{enumitem}
\usepackage{listings}
\usepackage{epstopdf}
\usepackage{array}
\usepackage{subcaption}
\usepackage{mathtools,cuted}
\usepackage{color}
\usepackage{colortbl}
\usepackage{cases}
\usepackage{float}

\usepackage{stackengine,graphicx}
\stackMath
\usepackage{amssymb}
\usepackage[normalem]{ulem}
\usepackage{amsmath,amssymb,amsfonts}
\usepackage{algorithmic}
\usepackage{graphicx}
\usepackage{textcomp}
\usepackage{xcolor}
\usepackage{subcaption}
\usepackage{bm}
\usepackage{mathrsfs}
\usepackage{verbatim}
\usepackage{comment}
\newcommand{\xbf}{\mathbf{x}}

\newcommand{\ebf}{\mathbf{e}}

\newcommand{\Etxt}{\textbf{E}}

\newcommand{\CC}{\mathbb{C}}
\def\BibTeX{{\rm B\kern-.05em{\sc i\kern-.025em b}\kern-.08em
    T\kern-.1667em\lower.7ex\hbox{E}\kern-.125emX}}

\makeatletter
\newcommand{\linebreakand}{%
  \end{@IEEEauthorhalign}
  \hfill\mbox{}\par
  \mbox{}\hfill\begin{@IEEEauthorhalign}
}
\makeatother

\begin{document}

\title{A Review of Electromagnetic Elimination Methods for low-field portable MRI scanner\\}

\author{

\small % Set font size to 10pt

\begin{tabular}[t]{c@{\extracolsep{8em}}c} 

1\textsuperscript{st} Wanyu Bian\textsuperscript{*} & 2\textsuperscript{nd} Panfeng Li \\
\textit{Harvard Medical School, Boston, USA} & \textit{University of Michigan, Ann Arbor, USA} \\
Corresponding author: wbian4@mgh.harvard.edu & \\

\\

3\textsuperscript{rd} Mengyao Zheng & 4\textsuperscript{th} Chihang Wang \\
\textit{Harvard University, Boston, USA} & \textit{New York University, New York, USA} \\

\\

5\textsuperscript{th} Anying Li & 6\textsuperscript{th} Ying Li \\
\textit{Xiaochang First People's Hospital, Xiaochang, China} & \textit{Xiaochang First People's Hospital, Xiaochang, China} \\

\\

7\textsuperscript{th} Haowei Ni & 8\textsuperscript{th} Zixuan Zeng \\
\textit{Columbia University, New York, USA} & \textit{Guangxi University, Nanning, China} \\

\end{tabular}
}

\maketitle

\begin{abstract}
This paper analyzes conventional and deep learning methods for eliminating electromagnetic interference (EMI) in MRI systems. We compare traditional analytical and adaptive techniques with advanced deep learning approaches. Key strengths and limitations of each method are highlighted. Recent advancements in active EMI elimination, such as external EMI receiver coils, are discussed alongside deep learning methods, which show superior EMI suppression by leveraging neural networks trained on MRI data. While deep learning improves EMI elimination and diagnostic capabilities, it introduces security and safety concerns, particularly in commercial applications. A balanced approach, integrating conventional reliability with deep learning’s advanced capabilities, is proposed for more effective EMI suppression in MRI systems.
\end{abstract}


\section{Introduction}
Ultra-low-field (ULF) portable MRI scanners, operating below 0.1T, offer a cost-effective alternative to traditional high-field MRI (1.5T or 3T) systems, which require expensive infrastructure. ULF MRI is particularly beneficial for underserved areas, allowing advanced imaging in locations such as rural clinics, emergency departments, and sports arenas. These portable scanners are especially effective for brain imaging, providing excellent soft tissue contrast for diagnosing neurological conditions. The development of ULF MRI expands access to diagnostic tools, easing the strain on traditional MRI facilities and improving healthcare efficiency.

Eliminating electromagnetic interference (EMI) is crucial for the effectiveness of ULF MRI scanners. EMI often from nearby electronic equipment, occurs at the Larmor frequency close to the operating frequency of MRI. This interference can disrupt MRI scan accuracy by inducing noise in the receiver coils. Traditional high-field MRI systems use bulky, enclosed RF shielding rooms (Faraday shield) to block this interference.

To make MRI technology more accessible and affordable, innovative EMI suppression strategies are essential. Techniques such as active shielding, adaptive filtering, and noise cancellation are being developed to mitigate EMI in ULF MRI systems. Active shielding creates counteracting electromagnetic fields, adaptive filtering adjusts dynamically to remove noise frequencies, and noise cancellation algorithms identify and subtract interference patterns.

These advanced EMI suppression methods enable the deployment of ULF MRI scanners in diverse settings without the need for expensive shielding infrastructure. This not only reduces costs but also expands access to high-quality MRI diagnostics for underserved populations.

This paper presents an analysis of both conventional and deep learning methods for EMI elimination without shielding rooms. We delved into a detailed comparison of these methods, highlighting the strengths and limitations of each. We explored the underlying principles of conventional analytical and adaptive EMI elimination techniques alongside cutting-edge deep learning approaches. 

\begin{figure*}[h]
\centering
{\includegraphics[width=0.7\paperwidth]{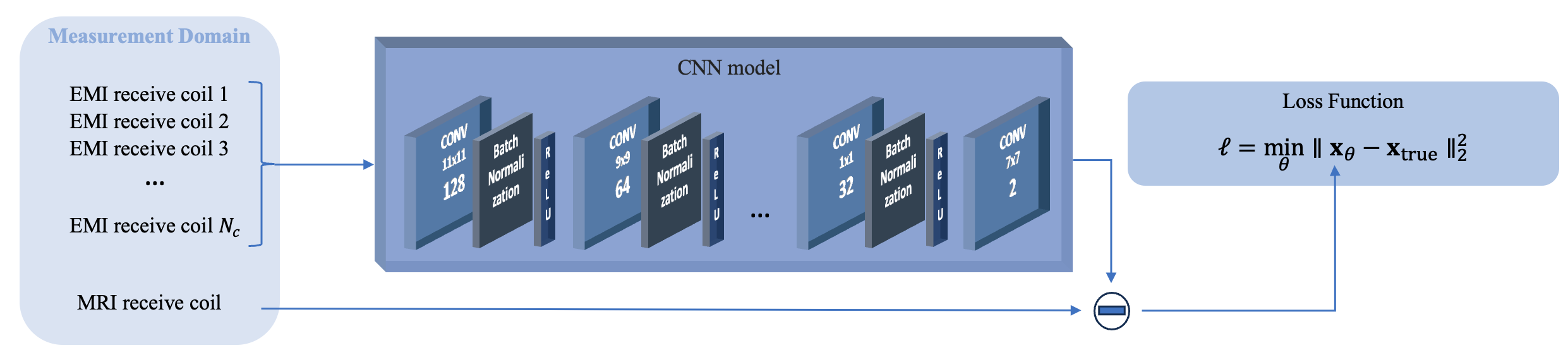}}
\caption{ General Network framework of deep learning EMI cancellation methods.}
\label{fig:CNN}
\end{figure*}

\section{Conventional Methods for EMI Reduction}
\label{sec:Conventional}
Recent advancements in active EMI elimination methods \cite{walsh2008multi,dyvorne2021freeing,srinivas2022external,yang2022active,parsa2023single,xu2024can} utilize  analytical and adaptive techniques to predict and mitigate EMI during MRI scans. Most of the methods involve the use of multiple external EMI receiver coils that are strategically positioned and oriented around the primary MRI receiver coil. These EMI receiver coils actively detect electromagnetic interference simultaneously with the primary MRI receiver coil~\cite{muller2020non,srinivas2020retrospective,xu2024llama}.

The primary MRI receiver coil is responsible for collecting the nuclear magnetic resonance (NMR) signals, which are crucial for imaging. However, it also collects interference signals and thermal noise, which can degrade the quality of the MRI images. The EMI receiver coils capture these unwanted EMI signals, providing valuable data to characterize the interference affecting the primary coil.

\begin{figure}[h]
\centering
\includegraphics[width=0.4\paperwidth]{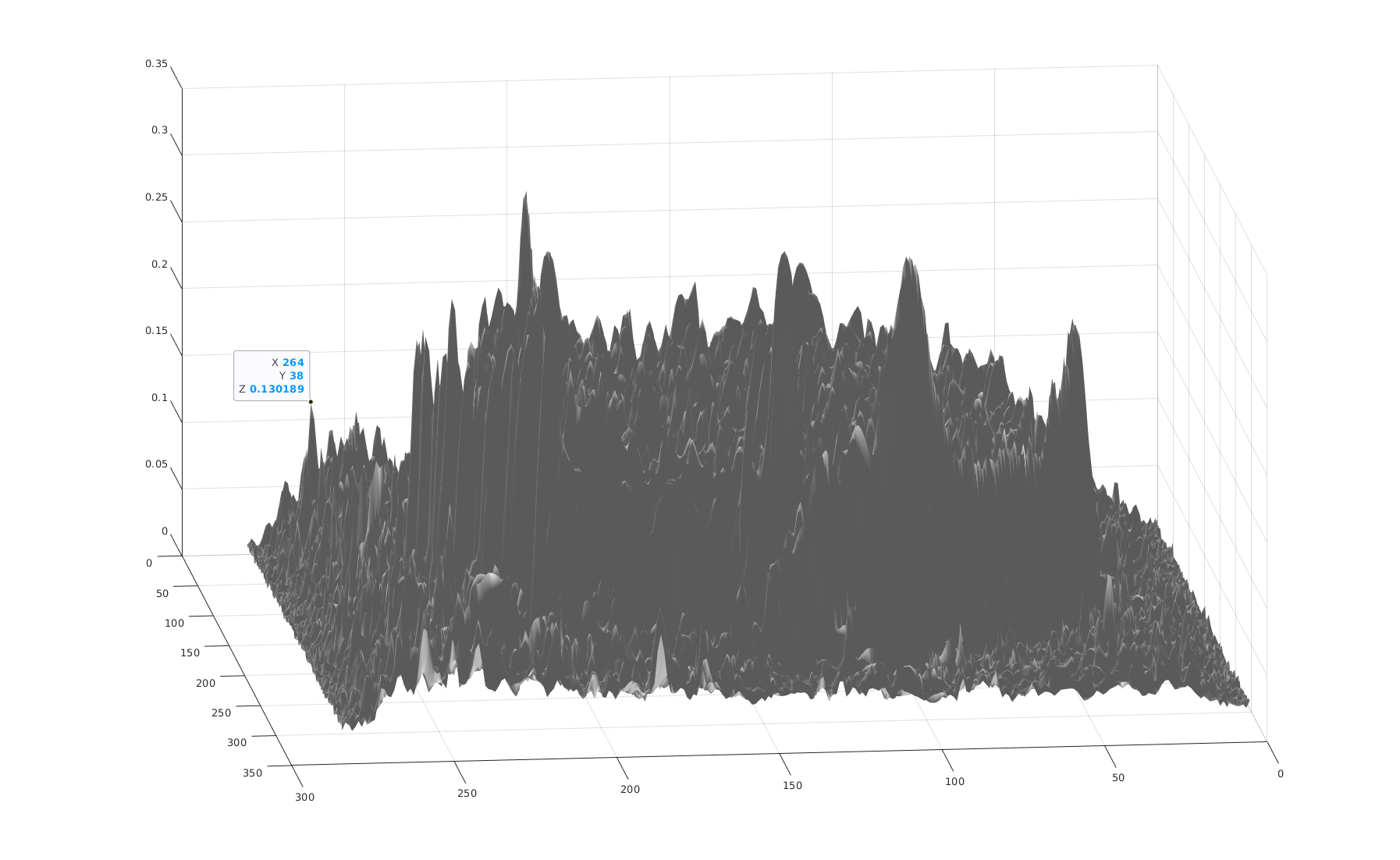}\\
\includegraphics[width=1\linewidth]{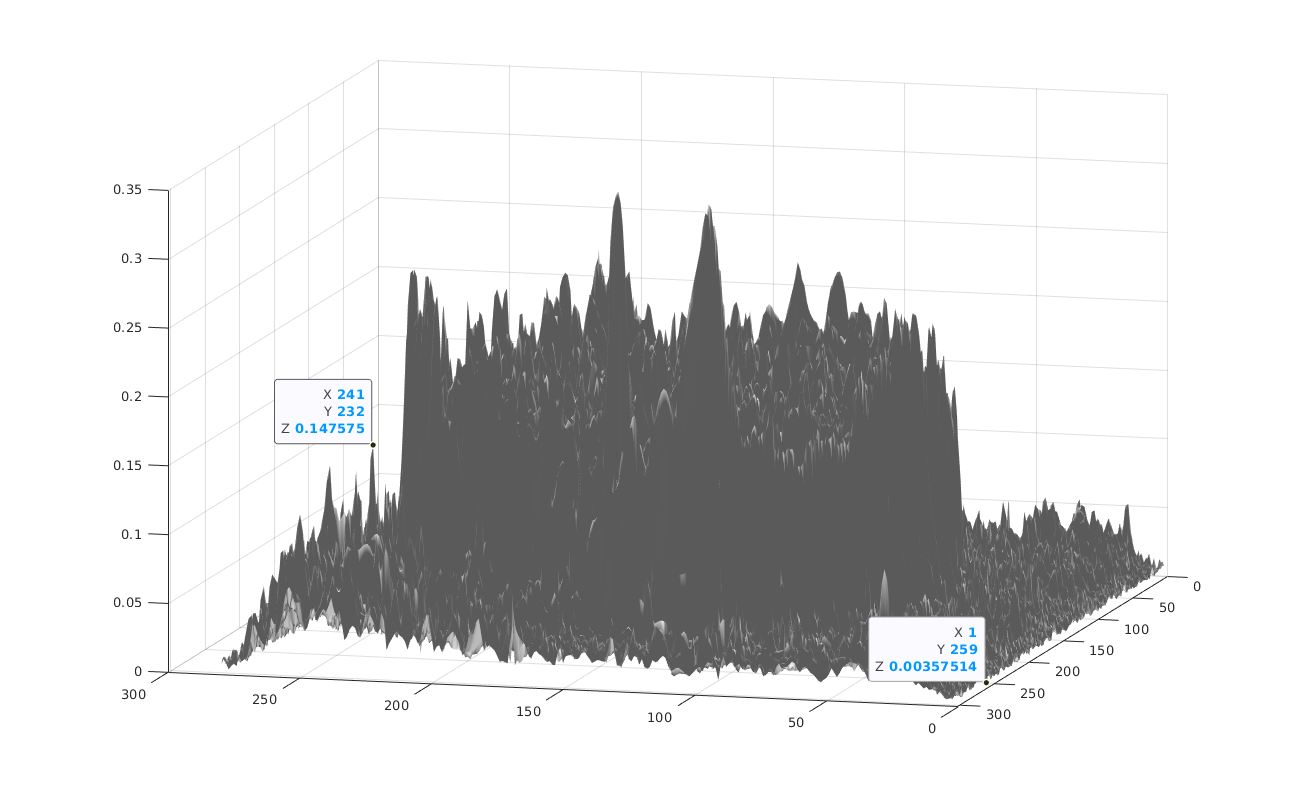}
\caption{ Absolute intensity of MRI signal and surrounding EMI signal.}
\label{fig:intensity}
\end{figure}

The captured EMI signals are then used to identify and characterize the unwanted noise present in the primary coil, the visual comparison between MR signal and EMI signals are illustrate in Fig \ref{fig:intensity}.
This characterized noise is processed retrospectively to cancel out the interference anomalies during the image reconstruction process. The formulations of the EMI in the primary coil employ frequency domain impulse response functions, which describe the relationship between the primary coil and the EMI receiver coils.
%By understanding these relationships, advanced algorithms can accurately model the interference and subtract it from the MRI data. This process significantly improves the quality of the resulting images by reducing noise and artifacts caused by EMI. These active EMI elimination methods are particularly beneficial in ULF MRI systems, where the sensitivity to external interference is higher due to the lower magnetic field strengths. Implementing such techniques allows ULF MRI scanners to achieve high-quality imaging without the need for bulky and expensive RF shielding.

\subsection{Multiple-external-coil-based Methods}
The raw measurement data captured from the primary coil $\xbf_0 \in \CC^{N_{x} \times N_{y}}$ is contaminated with EMI. This data comprises $N_x$ points in readout direction and $N_y$ points along phase encoding direction. The raw data $\xbf_0$ consists of the desired EMI-free measurement data $ \xbf^*_0 \in \CC^{N_{x} \times N_{y}}$ , and the EMI signal  $\ebf \in  \CC^{N_{x} \times N_{y}}$.
Assume there are  \( N_c \) external EMI receiver coils that detect the noise surrounding the primary coil. The set of EMI signals from these coils is denoted as \( \{ \xbf_i \mid i = 1, \ldots, N_c \} \). 

\subsubsection{EDITER}
EDITER \cite{srinivas2022external}  advances the management of time-varying EMI in MRI by dynamically adapting to changing EMI sources. It segments MRI data into smaller subsets based on empirical thresholds, allowing precise fitting of impulse response functions within limited temporal windows, such as individual or successive phase-encode lines.

By binning data and defining temporal windows, EDITER accurately estimates and eliminates EMI from each specific time frame, maintaining the integrity and quality of MRI images. The method captures EMI signals, fits them using impulse response functions, and subtracts them from the primary MR signal data to approximate true EMI-free signals.

EDITER algorithm utilizes these external EMI receiver coils to remove the noise and obtain the EMI-free signal by subtracting the EMI signal from the primary coil data, \( \xbf_0^* = \xbf_0 - \ebf \). However, since \( \ebf \) is unknown and difficult to measure directly, it is necessary to estimate $\ebf$  as accurately as possible by extracting representation information from the $N_c$ EMI receiver coils.

A noise characterization matrix $\Etxt$ was constructed  through a convolution operation in the time domain using all the external  EMI signal  $\{ \xbf_i| \, i = 1 \cdots N_c  \}$. There exists an impulse response vector $\vec{\gamma} $  such that $ \ebf \approx \Etxt  \vec{\gamma}$ .
To measure  $ \vec{\gamma} $, EDITER solves the optimization problem using measurement data $\xbf_0$ that obtained from primary coil:
\begin{equation}
 \min_{\vec{\gamma}} = || \xbf_0 - \Etxt  \vec{\gamma} ||^2_2.
\end{equation}
This yields:
\begin{equation}
 \vec{\gamma} = \Etxt^{\intercal} \xbf_0.
\end{equation}
The EMI free signal can be approximated as:
\begin{equation}
\xbf^*_0 \approx  \xbf_0 - \Etxt \vec{\gamma} = \xbf_0 -  \Etxt (\Etxt^{\intercal} \xbf_0).
\end{equation} This relationship works well under the situation that $\xbf_0 \approx \ebf $ because of the strong EMI signal in the primary coil.

\begin{table}
\caption{Performance on ULF EMI elimination}
\label{tab:perform_comparisons}
\begin{tabular}{@{}llll@{}}
\toprule
 \textbf{Model}            &       \textbf{Type}         & \textbf{ULF} & \textbf{Suppression rate}  \\ \midrule
 EDITER~\cite{srinivas2022external}   &  Conventional  & 0.055 \& 1.5 T & 95.7\% - 97.3\%  \\ \midrule
 AES System~\cite{yang2022active}     &  Conventional  & 0.05 T & 92.3\% \\ \midrule
 Liu et al.~\cite{liu2021low}       &  Deep Learning  & 0.055 T & 99.9\% \\ \midrule
 Zhao et al.~\cite{zhao2024electromagnetic} &  Deep Learning  & 0.055 \& 1.5 T &  99.9\%  \\ \bottomrule
\end{tabular}
\end{table}

Although achieving completely EMI-free images is challenging due to the complexity of electromagnetic interference, the EDITER method reduces EMI impact, enhancing overall image quality. 
To achieve improved image quality in in-vivo experiments, passive shielding remained necessary. Despite this, the EMI suppression performance of the receiving coil was not particularly effective. The system still required additional measures to adequately reduce electromagnetic interference.

\begin{table*}
\caption{Strengths and weaknesses of various ULF EMI elimination methods.}
\label{tab:comparisons}
\begin{tabular}{@{}lp{0.4\textwidth}p{0.4\textwidth}@{}}
\toprule
 \textbf{Model}           & \textbf{Strengths}                             & \textbf{Weakness} \\ \midrule
 EDITER~\cite{srinivas2022external}   & Dynamically adapts to changing EMI sources     & May still require passive shielding for in-vivo experiments \\ \midrule
 AES System~\cite{yang2022active}     & Decomposes data into frequency bands for targeted suppression     & Require synchronization of k-space data from multiple sources \\ \midrule
 Liu et al.~\cite{liu2021low}       & Outperforms traditional methods in robustness and simplicity     & Requires large amount of training data \\ \midrule
 Zhao et al.~\cite{zhao2024electromagnetic}  & Can model, predict, and remove complex EMI signals     & May introduce security vulnerabilities and safety risks \\ \bottomrule
\end{tabular}
\end{table*}

\subsubsection{Active EMI Suppression System}
EDITER offers a general concept for EMI suppression in ultra-low-field (ULF) portable MRI scanners, inspiring the development of various EMI suppression systems.

Lei Y et. al \cite{yang2022active} developed an active EMI suppression (AES) system tailored for a 50 mT portable MRI scanner without shielding. The primary source of interference was identified as human-body-coupled interferences. To address this, the authors created EMI detectors featuring a "ring"-shaped EMI receiving coil and an electrocardiograph electrode patch.
The AES system uses a reference channel-based algorithm to mitigate EMI in the MR signal captured by the RF receive coil. The method synchronizes the k-space data from the EMI detectors and the RF coil, decomposes this data into several frequency bands, and calculates transfer factors for each band. These factors are then applied to transform the reference EMI signals into corresponding interference signals, which are subtracted from the MR signal within each frequency band.

\subsection{Single-coil-based Method}
Parsa et al. \cite{parsa2023single} introduced an innovative single-coil-based technique to mitigate electromagnetic interference (EMI) in low-field MRI systems, building on the Halbach-based scanner without  a Faraday shield. The spectrometer has a built-in transmit/receive switch and preamplifier. A quadrature birdcage coil was chosen for its simplicity in achieving equal sensitivities in circularly polarized modes. In this configuration, the $B_0$-orthogonal (MR-active) channel captures both the MRI signal and EMI, while the $B_0$-parallel (MR-inactive) channel exclusively detects EMI. Although the birdcage coil itself is unshielded, the magnet's inner bore is equipped with a thin floating copper shield to reduce interactions between the gradient and RF coils.

The EMI reduction is achieved by combining signals from the MR-inactive and MR-active ports using a 180\degree power splitter/combiner. This combined output is then fed into the spectrometer's transmit/receive switch. This method is advantageous as it eliminates the need for external sensors or extensive signal processing, allowing implementation on a single receiver console. 
%However, when only a single receiver channel is available, the use of a 180\degree power splitter/combiner can reduce the signal-to-noise ratio (SNR) by approximately 40\%. This SNR loss can be avoided if the system has two separate receive channels.

\section{Deep Learning Methods for EMI Reduction}
\label{sec:DL}
Deep learning (DL) techniques are increasingly applied in MRI processing, showing remarkable effectiveness in tasks such as image reconstruction \cite{liang2020deep,knoll2020deep}, segmentation \cite{bernard2018deep}, and super-resolution \cite{pham2019multiscale}, etc. DL methods leverage large datasets and sophisticated algorithms to predict and mitigate EMI. These methods can learn complex patterns of interference and adaptively filter out noise, providing a more flexible and scalable approach to EMI elimination.

DL techniques \cite{zhao2022deep,liu2021low} employ neural networks trained on extensive MRI data, including both NMR signals and EMI signatures. By learning the characteristics of EMI, these models can accurately distinguish between true MRI signals and interference. 

One of the key advantages of  DL methods is their ability to handle non-linear and non-stationary EMI, common in real-world scenarios. These methods can be integrated into the image reconstruction pipeline, allowing for real-time or near-real-time EMI suppression. %This results in cleaner, higher-quality images without the need for extensive physical shielding.
%In addition, using  DL for EMI elimination can significantly reduce the overall cost and complexity of MRI systems. By relying on advanced software algorithms rather than hardware solutions, it becomes feasible to deploy high-quality MRI technology in a wider range of environments, including rural clinics and mobile health units. 

Most state-of-the-art DL methods for EMI reduction rely on supervised learning, which requires a large amount of ground truth data for training. This data is typically simulated from 3T scanners or collected in environments with RF shielding. However, obtaining sufficient EMI-free reference data is often impractical in real clinical settings due to the dynamic and variable nature of these environments.

Moreover, supervised learning methods are sensitive to domain shift, meaning they may not perform well when applied to different types of data than those they were trained on. This issue limits their effectiveness in diverse clinical settings where variations in equipment, patient populations, and environmental conditions are common.

\subsection{EMI Characterization Networks}
Y Liu et. al \cite{liu2021low}  developed a data-driven, deep learning-based model for EMI cancellation, which outperforms traditional analytical methods in robustness and simplicity for predicting and removing EMI.

EMI signals can change dynamically due to various surrounding sources and the influence of the human body, which acts as an effective antenna for EMI pickup. Body movement during scanning further affects these signals.  To address this, the system employs  ten EMI sensing coils  strategically placed EMI sensing coils around the scanner and inside the electronic cabinet to capture radiative EMI signals without RF shielding. These signals, along with those from the main MRI receive coil, were used to train a convolutional neural network (CNN).

The CNN model predicts the EMI component in the MRI receive coil signal for each frequency encoding (FE) line based on the EMI signals detected by the sensing coils. This predicted EMI component is then subtracted from the MRI receive coil signals, resulting in EMI-free k-space data before image reconstruction. Fig \ref{fig:CNN} illustrates the frametwork that using CNNs to characterize the representation of  EMI signals. %This deep learning EMI cancellation method effectively eliminates undesirable EMI signals, providing reliable results for both phantom and human brain imaging, even with dynamically changing environmental EMI sources. Experimental comparisons with a ground truth scenario, where an RF shielding cage was used to enclose the subject during scanning, demonstrated that this method nearly completely removes EMI noise without the shielding. The final image noise levels achieved with this method were as low as those obtained using the RF shielding cage, within a 5\% range. 

Y Zhao et. al \cite{zhao2024electromagnetic} presents an active sensing and deep learning approach for predicting and canceling  EMI in a low-cost  ULF 0.055 T permanent magnet brain MRI scanner, which operates without RF shielding and at a 2.32 MHz resonance frequency. Multiple EMI-sensing coils are placed around and inside the scanner to detect EMI signals during MRI scans. A five-layer CNN is trained to establish the relationship between EMI signals detected by the MRI receive coil and the EMI-sensing coils. The CNN model then predicts the EMI signal component detected by the MRI receive coil. This strategy effectively models, predicts, and removes EMI signals from MRI data using the scanner's multireceiver electronics and the physical properties of RF signal propagation, offering potential applications for other RF signal detection scenarios with complex EMI emissions.

\section{Benefits and Weaknesses of Conventioanal and Deep learning Methods}

Conventional methods for EMI suppression in MRI systems are advantageous due to their proven reliability and simplicity. These techniques have been extensively tested over the years, making them well-understood and trusted in the medical imaging community. They provide stable and reliable performance with relatively lower computational requirements, which translates to reduced costs and easier integration into existing MRI systems \cite{yu2024enhancing}. Conventional methods often involve straightforward implementation strategies, such as filtering and shielding, which are effective in many standard scenarios. However, their performance is limited in highly complex or dynamic EMI environments. They often lack the adaptability to quickly respond to new or unexpected sources of interference without significant manual adjustments or reconfiguration. %This inflexibility can result in suboptimal EMI suppression, especially in more challenging or evolving electromagnetic environments.

Deep learning methods offer superior performance by leveraging the power of large datasets and sophisticated algorithms. These methods can learn and adapt to various interference patterns, making them highly effective in distinguishing between genuine MRI signals and complex EMI. The automation provided by deep learning models significantly reduces the need for manual intervention, enhancing efficiency and consistency in EMI suppression. Furthermore, deep learning approaches can continuously improve over time as they are exposed to more data, potentially leading to incremental advancements in performance. However, these methods introduce potential security vulnerabilities and safety risks, particularly in production and commercial applications. The complexity of deep learning models can make them susceptible to adversarial attacks, where slight modifications to input data can lead to incorrect predictions. Additionally, ensuring the safety and robustness of these models in real-world scenarios requires extensive testing and validation. Deep learning methods also demand large amount of training data with reference true data, significant computational resources and expertise for their development, deployment, and maintenance, which can be a barrier to their widespread adoption, especially in resource-constrained settings. The strengths and weaknesses are summarized in Table \ref{tab:comparisons} and Table \ref{tab:perform_comparisons} shows the comparison of MEI noise suppression rates.

\section{Conclusion}
This paper presents a comprehensive analysis of both conventional and deep learning methods for EMI elimination in MRI systems, highlighting their respective strengths and limitations. While deep learning methods offer advanced performance in suppressing EMI by learning from large datasets and adapting to various interference patterns, they also bring challenges related to security, safety, and computational demands. Conventional methods, although less powerful in dynamic environments, provide a reliable and simpler alternative that is easier to implement and integrate into existing systems.
%The findings underscore the potential of deep learning to revolutionize EMI suppression in MRI technology, improving diagnostic accuracy and accessibility. %However, to fully realize this potential, it is crucial to address the security and safety concerns associated with deep learning, particularly in commercial and clinical applications. By balancing the strengths of both conventional and deep learning methods, the medical imaging community can develop more robust and effective strategies for EMI suppression, ensuring the continued advancement and reliability of MRI technology.

\renewcommand{\bibfont}{\footnotesize}

\footnotesize{
\bibliographystyle{IEEEtran}
\bibliography{main}

\begin{thebibliography}{10}
\providecommand{\url}[1]{#1}
\csname url@samestyle\endcsname
\providecommand{\newblock}{\relax}
\providecommand{\bibinfo}[2]{#2}
\providecommand{\BIBentrySTDinterwordspacing}{\spaceskip=0pt\relax}
\providecommand{\BIBentryALTinterwordstretchfactor}{4}
\providecommand{\BIBentryALTinterwordspacing}{\spaceskip=\fontdimen2\font plus
\BIBentryALTinterwordstretchfactor\fontdimen3\font minus \fontdimen4\font\relax}
\providecommand{\BIBforeignlanguage}[2]{{%
\expandafter\ifx\csname l@#1\endcsname\relax
\typeout{** WARNING: IEEEtran.bst: No hyphenation pattern has been}%
\typeout{** loaded for the language `#1'. Using the pattern for}%
\typeout{** the default language instead.}%
\else
\language=\csname l@#1\endcsname
\fi
#2}}
\providecommand{\BIBdecl}{\relax}
\BIBdecl

\bibitem{walsh2008multi}
D.~O. Walsh, ``Multi-channel surface nmr instrumentation and software for 1d/2d groundwater investigations,'' \emph{Applied Geophysics}, 2008.

\bibitem{dyvorne2021freeing}
H.~Dyvorne \emph{et~al.}, ``Freeing mri from its faraday cage with interference rejection,'' in \emph{ISMRM}, vol. 749, 2021.

\bibitem{srinivas2022external}
S.~A. Srinivas, ``External dynamic interference estimation and removal (editer) for low field mri,'' \emph{Magnetic resonance in medicine}, 2022.

\bibitem{yang2022active}
L.~Yang \emph{et~al.}, ``Active emi suppression system for a 50 mt unshielded portable mri scanner,'' \emph{IEEE Transactions TBME}, 2022.

\bibitem{parsa2023single}
J.~Parsa \emph{et~al.}, ``A single-coil-based method for electromagnetic interference reduction in point-of-care low field mri systems,'' \emph{Journal of Magnetic Resonance}, vol. 346, p. 107355, 2023.

\bibitem{xu2024can}
H.~Xu, J.~Ye, Y.~Li, and H.~Chen, ``Can speculative sampling accelerate react without compromising reasoning quality?'' in \emph{Tiny ICLR}, 2024.

\bibitem{muller2020non}
M.~M{\"u}ller-Petke, ``Non-remote reference noise cancellation-using reference data in the presence of surface-nmr signals,'' \emph{Geophysics}, 2020.

\bibitem{srinivas2020retrospective}
S.~A. Srinivas \emph{et~al.}, ``Retrospective electromagnetic interference mitigation in a portable low field mri system,'' \emph{ISMRM}, 2020.

\bibitem{xu2024llama}
H.~Xu, Y.~Li, and S.~Ji, ``Llamaf: An efficient llama2 architecture accelerator on embedded fpgas,'' \emph{arXiv preprint arXiv:2409.11424}, 2024.

\bibitem{liu2021low}
Y.~Liu \emph{et~al.}, ``A low-cost and shielding-free ultra-low-field brain mri scanner,'' \emph{Nature communications}, vol.~12, no.~1, p. 7238, 2021.

\bibitem{zhao2024electromagnetic}
Y.~Zhao \emph{et~al.}, ``Electromagnetic interference elimination via active sensing and deep learning prediction for radiofrequency shielding-free mri,'' \emph{NMR in Biomedicine}, vol.~37, no.~7, p. e4956, 2024.

\bibitem{liang2020deep}
D.~Liang \emph{et~al.}, ``Deep magnetic resonance image reconstruction: Inverse problems meet neural networks,'' \emph{Signal Processing Magazine}, 2020.

\bibitem{knoll2020deep}
F.~Knoll, ``Deep-learning methods for parallel magnetic resonance imaging reconstruction: A survey of the current approaches, trends, and issues,'' \emph{IEEE signal processing magazine}, vol.~37, pp. 128--140, 2020.

\bibitem{bernard2018deep}
O.~Bernard, ``Deep learning techniques for automatic mri cardiac multi-structures segmentation and diagnosis: is the problem solved?'' \emph{IEEE transactions on medical imaging}, vol.~37, no.~11, pp. 2514--2525, 2018.

\bibitem{pham2019multiscale}
C.-H. Pham, ``Multiscale brain mri super-resolution using deep 3d convolutional networks,'' \emph{Computerized Medical Imaging Graphics}, 2019.

\bibitem{zhao2022deep}
Y.~Zhao \emph{et~al.}, ``Deep learning driven emi prediction and elimination for rf shielding-free mri at 0.055 t and 1.5 t,'' in \emph{ISMRM}, 2022, p. 3864.

\bibitem{yu2024enhancing}
H.~Yu \emph{et~al.}, ``Enhancing healthcare through large language models: A study on medical question answering,'' \emph{arXiv:2408.04138}, 2024.

\end{thebibliography}
}

\end{document}